\documentclass[twocolumn,showpacs,amsmath,amssymb,prl]{revtex4}

\usepackage[dvips]{graphicx}

\newcommand{\figurewidth}{0.40\textwidth}
\newcommand{\smallerfigurewidth}{0.40\textwidth}

\begin{document}
\title{Polyelectrolyte Condensation Induced by Linear Cations}

\author{Camilo Gu\'aqueta}
\affiliation{%
    Department of Materials Science and Engineering,
    University of Illinois at Urbana-Champaign,
    Urbana, Illinois 61801
}
\author{Erik Luijten}
\email[Corresponding author. E-mail: ]{luijten@uiuc.edu}
\affiliation{%
    Department of Materials Science and Engineering,
    University of Illinois at Urbana-Champaign,
    Urbana, Illinois 61801
}

\date{\today}

\begin{abstract}
  We examine the role of the condensing agent in the formation of
  polyelectrolyte bundles, via grand-canonical Monte Carlo simulations.
  Following recent experiments we use linear, rigid divalent
  cations of various lengths to induce condensation. Our results clarify
  and explain the experimental results for short cations. For longer
  cations we observe novel condensation behavior owing to alignment of
  the cations. We also study the role of the polyelectrolyte surface
  charge density, and find a nonmonotonic variation in bundle stability.
  This nonmonotonicity captures two trends that have been observed in
  separate experiments.
\end{abstract}

\pacs{82.35.Rs, 87.15.Aa, 87.16.Ka}

\maketitle

Under conditions commonly found in biological systems, like-charged
polyelectrolytes in solution can form compact aggregates. This
phenomenon, referred to as polyelectrolyte condensation, occurs for a
variety of biopolymers, generally under the influence of charged
condensing agents such as multivalent ions~\cite{tang96,lyubartsev98},
short polyamines~\cite{wilson79}, or charged proteins~\cite{sanders05}.
If the polyelectrolytes are stiff chains, such as filamentous actin
(F-actin)~\cite{tang96,sanders05} or the fd and M13
virus~\cite{lyubartsev98}, they aggregate into dense, hexagonally
coordinated bundles. The counterintuitive nature of the effective
attraction implied by bundle formation, in combination with the
biological importance of the phenomenon, have made it the topic of
numerous studies over the past decades (see
Refs.~\cite{ray94,gronbech-jensen97,ha97,solis99,shklovskii99a,naji04}
and references therein). There is now general agreement that strong
electrostatic correlations are a crucial
condition~\cite{shklovskii99a,naji05}, in accordance with the
experimental observation that most condensing agents carry a multivalent
charge.  However, other factors that affect the tendency of a system to
exhibit condensation are much less well established, mainly because
these factors are difficult to disentangle in both experiments and
theory.

Even for a simple ionic condensing agent, its interaction strength with
the polyelectrolyte is affected not only by its valency, but also by its
size~\cite{solis01} and by the surface charge density of the
polyelectrolyte. In addition, the ionic concentration plays an important
role, as it determines the stabilizing osmotic pressure exerted on the
bundle by the surrounding solution~\cite{sanders05,lyubartsev98} and
also controls the strength of entropic effects such as counterion
release and depletion interactions~\cite{ray94}. Thus, even a systematic
variation of the ion size can yield results that are difficult to
interpret, as it alters both the binding of the ion to the
polyelectrolyte and the osmotic pressure of the solution.  Likewise,
variation of the polyelectrolyte charge not only affects ionic binding,
but also the direct electrostatic repulsion between polyelectrolytes.
Accordingly, only an integrated approach can resolve the true origin of
observed trends in the aggregation behavior.

There is a remarkable dearth of such studies.  Experimentally, there are
technical limitations. For example, examination of the effect of
counterion size is hampered by the uncertainty in measuring hydrated ion
sizes~\cite{tang02} and variation of counterion valency often entails
the simultaneous variation of other ionic properties.  Furthermore, it
is difficult to measure the ionic concentration within the aggregate and
hence to assess the osmotic effects arising from a concentration
imbalance with the bulk solution.  This osmotic stress is also often
ignored in computational and theoretical studies.  Inspired by two
recent experimental studies~\cite{tang02,butler03}, in this Letter we
aim to obtain a more complete understanding of how bundle formation is
affected by (i)~the size of the condensing agent and (ii)~the surface
charge density of the polyelectrolyte.  In Ref.~\cite{butler03}, the M13
virus was bundled using diamine molecules.  In solution the diamines
form divalent cations
with a length that can be systematically varied. It was found that only
the shortest diamines can induce bundle formation. This implies that
increasing the diamine size decreases bundle stability, although a
detailed analysis of the role of diamine size could not be obtained.
Furthermore, it was found that increasing the M13 surface charge
density~$\sigma$ also destabilizes the bundle. In contrast, in
Ref.~\cite{tang02} this virus was bundled using alkali earth metal ions
and an \emph{increased} stability was observed upon increase of the
surface charge density~\cite{note-fd}.  It is our purpose to clarify
these experimental findings through computer simulations.  Indeed,
earlier simulations~\cite{tang02} confirmed part of the observations,
but did not explain the surface-charge dependence observed in
Ref.~\cite{butler03} and did not address effects arising from internal
degrees of freedom of the condensing agent.  We confirm that an increase
of the diamine size destabilizes the bundle, but also demonstrate how
stability is recovered for even longer diamines through changes in their
spatial arrangement.  Furthermore, we reconcile the contradictory
experimental findings for the effect of surface charge density. As the
underlying mechanisms are highly generic, our findings are relevant for
broad classes of systems that display electrostatically induced
aggregation.

We employ Monte Carlo simulations of a model based upon the experimental
systems~\cite{tang02,butler03}. M13 is modeled as an infinitely long
cylindrical rod, with monovalent charges placed on a rectangular grid
wrapped around the rod at a radial distance of $28$~\AA\@. The surface
charge density is controlled via the lattice parameters of the
grid. 
Soft repulsive $1/r^{12}$ interactions between the cylinder and other
particles bring the effective cylinder radius to $R=30.5$~\AA\@.  The
diamines are modeled as rigid straight molecules composed of soft
repulsive beads of effective diameter $d=5$~\AA. The terminal beads,
both carrying a monovalent charge, have a center-to-center
separation~$\delta$ and are connected by $\lceil \delta/d \rceil - 1$
regularly spaced uncharged beads. The coions are monovalent beads of the
same diameter.  Water is represented as a homogeneous dielectric medium
($\varepsilon=80$) and the temperature is set to $T=298$~K.
Electrostatic interactions are calculated via Ewald summation. The
polyelectrolytes are placed in a periodic cell, forming an infinite
hexagonal array with center-to-center rod-rod separation $L$.  To reduce
finite-size effects a $2\times 2$ array of rods is used. Rod degrees of
freedom are ignored. As in Refs.~\cite{lyubartsev98,tang02}, we employ
the grand-canonical ensemble to ensure that the bundle is in chemical
equilibrium with a bulk solution of diamine salt (diamines plus coions).
Thus, in addition to a fixed number (300--500) of diamines that balance
the polyelectrolyte charge, the bundle contains a fluctuating amount of
diamine salt. We perform separate simulations of the bulk solution to
establish its osmotic pressure~$\Pi_{\textrm{bulk}}$ as a function of
chemical potential.  Mechanical
equilibrium 
is then obtained if the net osmotic pressure
$\Pi=(\Pi_{\textrm{bundle}}-\Pi_{\textrm{bulk}})$ 
vanishes.

\begin{figure}
  \centerline{\includegraphics[width=\figurewidth]{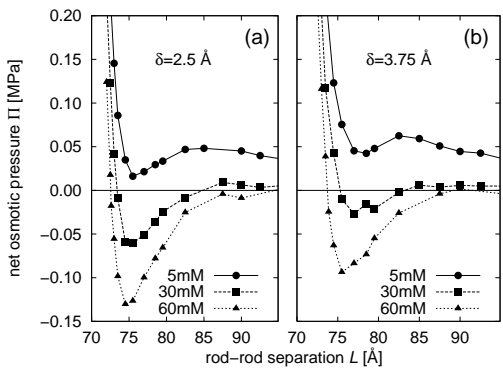}}
  \caption{Net osmotic pressure as a function of rod-rod separation~$L$,
  for bulk diamine concentrations of 5, 30, and 60~mM, at a
  polyelectrolyte surface charge density $\sigma=0.633$~$e$/nm$^2$. The
  diamine charge separation is (a)~$\delta=2.5$~\AA, and
  (b)~$\delta=3.75$~\AA\@. Error bars are comparable to the symbol
  size.}
  \label{fig:maxwell}
\end{figure}

Figure~\ref{fig:maxwell}(a) shows the net osmotic pressure as a function
of rod-rod separation, for diamines with length $\delta=2.5$~\AA, at
three bulk diamine concentrations.  Negative $\Pi$ corresponds to bundle
contraction; a zero-crossing at small $L$ indicates a free-energy
minimum and yields the rod separation of a stable bundle. At low diamine
concentrations $\Pi$ is always positive and no condensation takes place,
but as the concentration exceeds a threshold value the pressure crosses
zero and bundles form. This agrees with experimental results on
M13 and fd~\cite{tang02,butler03}, and highlights the importance of the
excess bulk diamine concentration.
Experimentally~\cite{butler03}, it is found that the bundles become
unstable when $\delta$ increases.  This is also reproduced by our
simulations: for $\delta = 3.75$~\AA\ [Fig.~\ref{fig:maxwell}(b)] the
osmotic pressure curves are shifted outwards and upwards, reflecting a
decrease in stability.  To study this in detail, we vary the diamine
length over a much wider range, $2.5\textrm{ \AA} \leq \delta \leq
22.5\textrm{ \AA}$.  Surprisingly, two distinct regimes emerge.  The
trend observed experimentally and pictured in Fig.~\ref{fig:maxwell}
continues up to $\delta \approx 7.5$~\AA\@, but for larger diamine
lengths---not studied in Ref.~\cite{butler03}---this trend
\emph{reverses} and bundle stability \emph{increases} with $\delta$.

\begin{figure}
  \centerline{\includegraphics[width=\figurewidth]{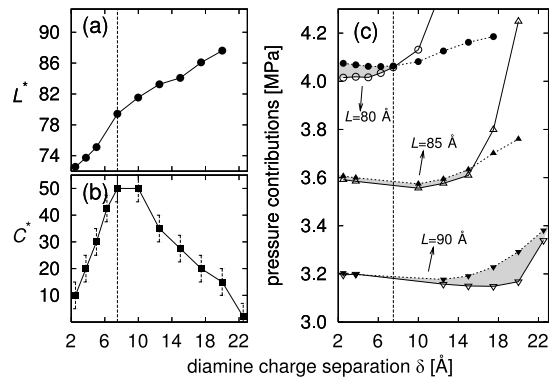}}
  \caption{Effect of diamine charge separation on bundle stability for
  polyelectrolytes with $\sigma=0.633$~$e$/nm$^2$. (a) Stable rod-rod
  separation at fixed 60~mM diamine concentration. (b) Bulk diamine
  concentration~$C^*$ at which bundling first occurs. (c) Separate
  contributions to the net osmotic pressure, for $L=80$, $85$, and
  $90$~\AA, at fixed 60~mM diamine concentration. The dashed curves give
  the negative of the electrostatic contribution (i.e., electrostatic
  \emph{attraction}), and the solid curves give the short-range
  contribution to the pressure.}
  \label{fig:llength}
\end{figure}

To quantify and understand this behavior we refer to
Fig.~\ref{fig:llength}.  The stable rod separation~$L^*$ (at fixed,
sufficiently high diamine concentration) is shown \emph{vs.}\ diamine
length in Fig.~\ref{fig:llength}(a), and the threshold diamine
concentration~$C^*$ required for condensation~\cite{note-reweight} in
Fig.~\ref{fig:llength}(b). In both panels, a vertical line separates the
two regimes. For smaller diamines, $\delta \lesssim 7.5$~\AA, the bundle
swells linearly with $\delta$.  This is accompanied by an increase in
$C^*$, i.e., a larger bulk osmotic pressure is needed to maintain bundle
stability. To understand this decrease in stability it is instructive to
consider how the partitioning of diamines between the bundle and the
bulk changes when $\delta$ increases~\cite{sanders05}. On the one hand,
larger diamines are more likely to be excluded from the bundle, thus
enhancing the stabilizing effect of the bulk solution. On the other
hand, swelling allows more diamine salt to enter the bundle, decreasing
the concentration difference. 
We find that the latter effect dominates: as $\delta$ increases from
$2.5$~\AA\ to $7.5$~\AA\ the diamine salt concentration of in the bundle
almost doubles.  Accordingly, the bulk solution becomes less effective
at holding the bundle together.

What drives the swelling of the bundle? Wong \emph{et
al.}~\cite{butler03} propose that longer diamines behave effectively as
two monovalent ions rather than a single divalent unit, with the
consequent loss of electrostatic attraction leading to dissolution of
the bundle. We examine this idea in Fig.~\ref{fig:llength}(c), which
shows how the separate contributions to the net osmotic pressure vary
with diamine size. 
Three different rod separations are plotted, spanning the range of
Fig.~\ref{fig:llength}(a). The solid curves (open symbols) show the
short-range contribution to the pressure, which includes the kinetic and
excluded-volume terms, and the dashed curves (filled symbols) show the
\emph{negative} of the electrostatic contribution. Thus, the
intersection of two curves corresponds to $\Pi=0$, i.e., a stable bundle
for the given $\delta$, and in the shaded regions $\Pi$ is negative.
The top pair of curves ($L=80$~\AA) is typical for the regime $\delta
\lesssim 7.5$~\AA\@.  Rather than the suggested loss of electrostatic
attraction~\cite{butler03} we see that the electrostatic contribution is
relatively \emph{constant}, and it is instead the short-range
\emph{repulsion} that changes, rising sharply with $\delta$ and causing
the bundle to swell and lose stability.

\begin{figure}
  \centerline{\includegraphics[width=\figurewidth]{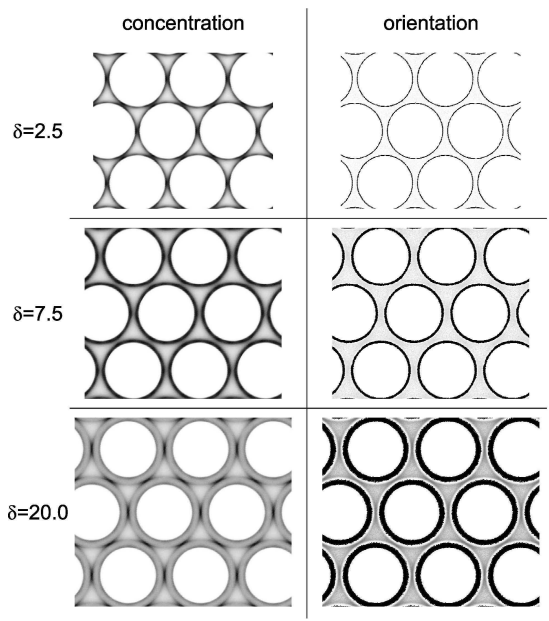}}
  \caption{Fluid structure within stable bundles, projected onto a plane
  perpendicular to the bundle axis, for three different diamine lengths
  (top to bottom: $\delta=2.5$, $7.5$, and $20$~\AA).  Left: diamine
  concentration; right: diamine orientation with respect to
  polyelectrolyte axis. For discussion see the text.}
  \label{fig:NS}
\end{figure}

We now turn to the regime of large $\delta$. Near $\delta = 7.5$~\AA\
there is a weak change in slope in Fig.~\ref{fig:llength}(a), indicating
a slower swelling of the bundle. More importantly,
the threshold diamine concentration starts to \emph{decrease}
[Fig.~\ref{fig:llength}(b)]. This reversal of $C^*$ is particularly
remarkable in view of the continued swelling of the bundle, as it
implies that bundle stability increases \emph{despite} the weakening of
the salt imbalance. Indeed, Fig.~\ref{fig:llength}(c) shows that at
large $\delta$ the electrostatic attraction no longer remains constant,
but instead rises. 
Important insights into this phenomenon can be obtained from the
distribution and orientation of diamines in the bundle, as shown in
Fig.~\ref{fig:NS} for fixed bulk concentration. The three rows
correspond to $\delta=2.5$, $7.5$, and $20$~\AA, each shown at the
corresponding stable rod-rod separation.  The left-hand column shows the
spatial distribution of diamines in a cross-section of the bundle, with
darker shading corresponding to higher concentration. The right-hand
column shows how the diamines are oriented with respect to the rods,
with black indicating a tendency to align parallel to the rods
(out-of-plane), grey indicating perpendicular alignment (in-plane), and
white indicating isotropic orientation. For small diamines
($\delta=2.5$~\AA) the bundle is quite compact, and the diamines are
concentrated mostly in the two-fold bridging sites between pairs of
rods. The orientation of the diamines is essentially isotropic.  At
$\delta=7.5$~\AA---where the bundle is swollen and has a low
stability---the diamines are much less concentrated in the bridging
sites but rather form a condensed layer around each polyelectrolyte.
The diamines remain mostly isotropic, but have a slight tendency to
align parallel to the rods in the condensed layer and perpendicular to
the rods elsewhere. As the diamine length increases further
($\delta=20$~\AA), however, a more complex structure appears, where the
diamines are once again strongly concentrated in the bridging sites but
\emph{also} exhibit a high degree of alignment. Close to the rods the
diamines orient themselves in parallel, and in the bridging sites they
orient themselves perpendicular to the rods. We have explicitly verified
that in the bridging sites the diamines are oriented such that both
monovalent ends lie within the condensed layer of neighboring rods.
Thus, longer diamines act as ``linkers'' between the rods, which
accounts for the rise in electrostatic attraction and the resurgence of
bundle stability.
  
Lastly, we consider the relationship between polyelectrolyte surface
charge density~$\sigma$ and bundle stability. Increasing $\sigma$
enhances the direct rod repulsion as well as the coupling between the
rod and the condensing agent. Theoretical work~\cite{naji04,naji05}
predicts that the latter effect dominates: an effective attraction
occurs when both the coupling parameter
$\Xi=q^2 \ell_{\textrm{B}}/\mu$ and the Manning parameter
$\xi=R/\mu$ are sufficiently large, where $q=2$ is the diamine valency,
$\ell_{\textrm{B}} = e^2/(4\pi \varepsilon\varepsilon_0
k_{\textrm{B}}T)$ the Bjerrum length, and $\mu=(2\pi q \ell_{\textrm{B}}
\sigma)^{-1}$ the Gouy-Chapman length. Both $\Xi$ and $\xi$ increase
with $\sigma$, so that the net electrostatic attraction is enhanced at
larger surface charge densities. There is another consideration,
however: more highly charged rods require a higher concentration of
diamines to maintain electroneutrality of the bundle, leading to an
increase in the repulsive short-range contribution to the pressure. A
priori, it is not clear which of these trends---enhanced electrostatic
attractions, or larger excluded-volume repulsions---dominates, and there
is experimental evidence supporting both scenarios. With diamines as the
condensing agent, an increase in surface charge density from
$0.303$~$e$/nm$^2$ to $0.343$~$e$/nm$^2$ dissolves the bundle, and a
large increase in diamine concentration is needed to regain bundle
stability~\cite{butler03}. But for certain divalent cations an increase
in $\sigma$ from $0.343$~$e$/nm$^2$ to
$0.457$~$e$/nm$^2$~\cite{note-sigma} can enhance bundle
stability~\cite{tang02}. Thus, the experimentally observed trend appears
to depend on the condensing agent.

\begin{figure}
  \centerline{\includegraphics[width=\smallerfigurewidth]{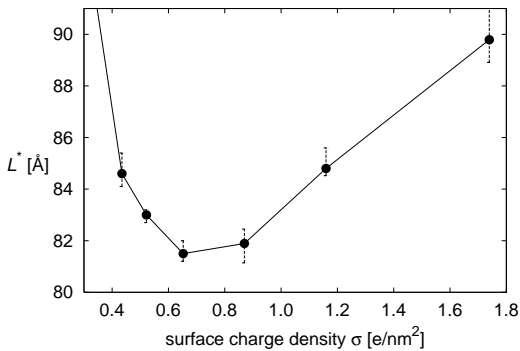}}
  \caption{Stable rod separation~$L^*$ as a function of polyelectrolyte
  surface charge density~$\sigma$, for $\delta = 10$~\AA\ at 60~mM bulk
  diamine concentration. The curve
  demonstrates that increasing $\sigma$ can either increase or decrease
  bundle stability.}
  \label{fig:sigma}
\end{figure}

We shed some light on this situation by varying $\sigma$ in our
simulations from $0.326$~$e$/nm$^2$ to $1.740$~$e$/nm$^2$, at constant
diamine length $\delta=10$~\AA\@. For this $\delta$ the degree of
alignment within the bundle is relatively low and partitioning of
diamines between bundle and bulk is quite important, so variation of
$L^*$ is a strong indication of bundle stability. At a fixed diamine
concentration of 60~mM, $L^*$ indeed shows a strikingly nonmonotonic
variation with $\sigma$ (Fig.~\ref{fig:sigma}). For small surface charge
density the bundle stability increases with $\sigma$, as expected from
the enhanced electrostatic coupling, yet for $\sigma \gtrsim
0.7$~$e$/nm$^2$ this trend is \emph{reversed}, owing to the large number
of neutralizing diamines within the bundle.  This reversal suggests that
by varying $\sigma$ over a sufficiently wide range, it may be possible
to observe in a single set of experiments, with a single condensing
agent, both of the trends which have previously been observed only
separately.

It is worthwhile to comment on the five dimensionless parameters that
characterize the system~\cite{note-r-eff}.  The coupling parameter~$\Xi$
and the Manning parameter $\xi$ are convenient choices for two of these.
Two further parameters account for the role of rod-rod separation and
bulk diamine concentration. To account for diamine size,
Ref.~\cite{butler03} proposed $\psi \equiv \mu/\delta$, with the
suggestion that $\psi>1$ is necessary for bundle formation.  However,
our demonstration that bundle stability is nonmonotonic in $\delta$
makes this choice of $\psi$ problematic.  Moreover, the criterion
$\mu/\delta>1$ predicts a dependence on temperature, valency, and
dielectric constant that contradicts well-established trends for bundle
stability. Instead, we suggest the parameter $\delta/a_s$, where $a_s$
is the charge separation on the polyelectrolyte surface. This choice was
considered in Ref.~\cite{butler03}, but disregarded because it did not
vary significantly in the experiments.  Here, we have varied
$\delta/a_s$ over a much larger range. Interestingly, if this is done
through variation of $\sigma$ (Fig.~\ref{fig:sigma}), the most stable
condition indeed corresponds to $\delta/a_s = \mathcal{O}(1)$, i.e., the
diamine charge separation matches the charge separation on the rod
surface. However, we also note that this parameter cannot capture the
nonmonotonicity observed as a function of $\delta$
[Fig.~\ref{fig:llength}(b)], as it does not account for the alignment
effects occurring for long diamine molecules.

In summary, we have clarified several mechanisms governing the role of
condensing agents in polyelectrolyte bundling. We have also presented a
unified picture for the dependence on surface charge density, combining
seemingly conflicting experimental observations.

\begin{acknowledgments}
  This material is based upon work supported by the National Science
  Foundation under CAREER Grant No.\ DMR-0346914 and Grant No.\
  CTS-0120978 via the WaterCAMPWS Science and Technology Center.
\end{acknowledgments}


\begin{thebibliography}{10}
\bibitem{note-fd}
In Ref.~\cite{tang02}, variation of the surface charge was realized by
  exchanging M13 for the structurally identical fd virus.
\bibitem{note-reweight}
For each $\delta$ the threshold concentration is obtained through interpolation
  of simulation results for concentrations bracketing~$C^*$, using histogram
  reweighting.
\bibitem{note-sigma}
These surface charge densities are calculated based upon the value reported in
  Ref.~\cite{butler03} and the difference between M13 and fd.
\bibitem{note-r-eff}
We assume that the diameter of the diamine beads can be absorbed in a
  redefinition of the rod radius~\cite{naji05}.
\end{thebibliography}

\end{document}